# A QUANTUM KEY DISTRIBUTION NETWORK THROUGH SINGLE MODE OPTICAL FIBER

Muhammad Mubashir Khan[1], Salahuddin Hyder[2], Mahmood K Pathan[3], Kashif H Sheikh[4]
*Department of Computer Science & IT, NED University of Engg: & Tech, Karachi, Pakistan*

*Email (mmkhan@comp.leeds.ac.uk[1], hyder@cs.mcgill.ca[2], mkpathan@neduet.edu.pk[3], skashifh@yahoo.com[4])*

## ABSTRACT

*Quantum key distribution (QKD) has been developed within the last decade that is provably secure against arbitrary computing power, and even against quantum computer attacks. Now there is a strong need of research to exploit this technology in the existing communication networks. In this paper we have presented various experimental results pertaining to QKD like Raw key rate and Quantum bit error rate (QBER). We found these results over 25 km single mode optical fiber. The experimental setup implemented the enhanced version of BB84 QKD protocol. Based upon the results obtained, we have presented a network design which can be implemented for the realization of large scale QKD networks. Furthermore, several new ideas are presented and discussed to integrate the QKD technique in the classical communication networks.*

## 1. INTRODUCTION

Quantum Key Distribution (QKD) is a technique which has been developed for securing data transmission by means of quantum mechanical rules [1]. QKD was first proposed by Bennett & Brassard in 1984 [2]. But owing to some technological challenges it could not get practical realization. In the beginning of 21st century two companies of the world one from USA, MagiQ Tech, and another from Switzerland, idQuantique, proposed the practical implementation of quantum cryptography. The practical realization of QKD has opened new directions of research in the area of quantum cryptography. Different types of attacks on QKD systems and the techniques for their prevention have been proposed by the research community [3]. Several issues like error-correction and privacy amplification for the QKD systems have also been raised. At the time of writing this paper QKD is assumed to be more protected than any other known cryptosystem against classical as well as quantum attacks. We tested the QKD using 25 km single mode optical fiber and calculated the important parameters related to the security of the QKD cryptosystem. In this paper we have presented our experimental set up and the results obtained.

Another hot issue of research is to exploit the QKD technology in the existing networks to achieve highest degree of security. The purpose of practically realizing the QKD is to find ways to establish a QKD network. In this regard the main concern is to integrate the QKD in the existing communication infrastructure. In this paper, we have presented a new model of QKD network which can be implemented without using any far-reaching technology which is not present today. Several issues about the improvement and practical implementation of this model are also discussed.

## 2. OUR QKD EXPERIMENT

We designed a simple point-to-point QKD network by using the QKD emitter and receiver from Swiss company idQuantique. The two panels of emitter and receiver are conventionally named as Alice and Bob respectively. Figure 1.

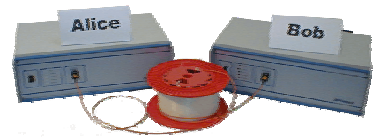

**Figure 1. QKD Emitter & Receiver From idQuantique**

There are two communication channels between Alice and Bob. One is classical communication channel and the other is quantum communication channel. The two panels are directly connected to each other through quantum communication channel which is a single mode optical fiber. We use single mode optical fiber because of the fact that in multimode optical fiber the modes couple easily, acting on the qubit like a non-isolated environment [1]. Hence multimode fibers are not appropriate as quantum channels. The quantum communication channel is capable of transmitting photons from Alice to Bob. Alice and Bob are indirectly connected to each other through two computers which are directly connected to each other through classical communication link. The whole network design is shown in figure 2.

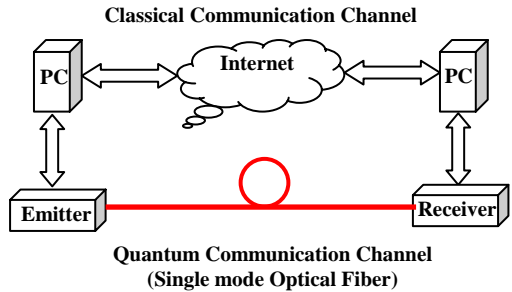

**Figure 2. Simple Point-to-Point QKD Network**

### 2.1. Specifics of our QKD Test Bed

The principle of the QKD auto compensating setup which is explained in detail in [1], is such that the key is encoded in the phase between two pulses traveling from Alice to Bob and back, see figure 3. A strong laser pulse at the rate of 1550 nm emitted at Bob is separated at a 50/50 beam splitter. The two pulses impinge on the input ports of a polarization beam splitter (PBS), after having traveled through a long arm and a short arm, including a phase modulator and a 50ns delay line, respectively. All fibers and optical elements at Bob are polarization maintaining. The linear polarization is turned by $90°$ in the short arm therefore the two pulses exit Bob's setup by the same port of PBS. The pulses travel down to Alice, are reflected on a Faraday Mirror, attenuated and come back orthogonally polarized. In turn both pulses now take the other path at bob and arrive at the same time at beam splitter where they interfere. They are then detected either in one detector (D1), or after passing through the circulator in another detector (D2). Since the two pulses take the same path, inside Bob in a reversed order, this interferometer is auto-compensated. To implement the BB84 protocol, Alice applies a phase shift of ($\phi_A = 0$ or $\pi$) and ($\phi_A = \pi/2$ or $3\pi/2$) on the second pulse with is phase modulator $PM_A$. Bob chooses the measurement basis by applying ($\phi_B = 0$ or $\pi/2$) shift on the first pulse on its way back. The photon counters are Peltier-cooled, actively gated, InGaS/InP APDs [10]. The dark count noise of the detectors is around $10^{-5}$ per gate. The variable attenuator (VA) at Alice is set to a low level and bright laser pulses are emitted by Bob. The time delay between the triggering of the laser and a train of gates of the detectors is scanned until the reflected pulses are detected. The delays between the the two 2.5ns detection gates are adjusted, as well as the timing of the 50 ns pulse applied on the phase modulator $PM_B$. For a storage line measuring approximately 10 km, a pulse train contains 480 pulses at a frequency of 5 $MH_z$. A 90% coupler ($BS_{10/90}$) directs most of the incoming light pulses to an APD-detector module ($D_A$). It generates the trigger signal used to synchronize the Alice's 20 $MH_z$ clock with the one of Bob. This synchronized clock allows Alice to apply a 50 ns pulse at the phase modulator $PM_A$ exactly when the second weak pulse passes. Only this second pulse contains the phase information and must be attenuated below the one photon per pulse level. Further details may be requested from idQuantique.

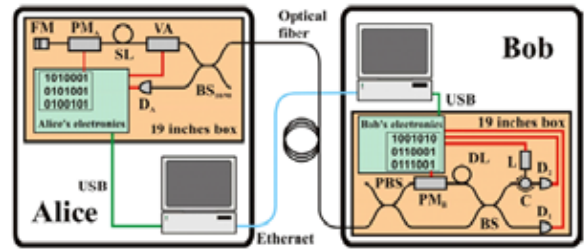

**Figure 3. Schematic of QKD Prototype**

### 2.2. Experimental Results

We tested the QKD network in the laboratory using the 25 Km single mode optical fiber spool featuring the losses of approximately 11 dB. The average number of photons per pulse was taken as $\mu = 0.1$. As the raw key creation rate ($R_{raw}$), see [7] for its detailed description, is one of the important parameters to characterize the performance of QKD systems. We calculated $R_{raw}$ with the following formula.

$$R_{raw} = q\mu\nu\eta_t\eta_d \qquad (1)$$

where $q$ is the systematic factor, which is 0.5 for four state BB84 protocol, $\mu$ is the average number of photons per pulse, $\nu$ is the repetition frequency which is 5 MH$_Z$ in our case, $\eta_d$ is the quantum detection efficiency and $\eta_t$ is the transmission efficiency. We found the raw key rate of about 490Hz.

The quantum bit error rate (QBER) which is also an important parameter to characterize the QKD system was found to be 4.5%. Generally the QBER is given as

$$QBER = \frac{false\_counts}{total\_counts} = \frac{false\_counts}{false\_counts + correct\_counts}$$

The purpose of our experiment is to practically analyze the feasibility of this QKD setup to design the specific QKD network in which this QKD technology can be practically implemented with minor modifications. Several international research groups have tested the QKD with varying successful results, as shown in Table 1.

**Table 1. Experimental Realization of QKD by Different Groups**

| Group | Distance [km] | $\mu$ | $R_{raw}$[Hz] | Measured QBER |
|---|---|---|---|---|
| Geneva | 22.8 | 0.1 | 486 | 4.5 % |
| BT | 25 | 0.15 | 500 | 2 % |
| Los Alamos | 24 | 0.4 | 20 | 1.6 % |

Today many companies are investing in quantum-cryptography systems. IBM's Almaden Research Center, the NEC Research Institute, Toshiba, and Hewlett-Packard are on the brink of introducing products. In March 2004, NEC scientists in Japan sent a single photon over a 150-km fiber-optic link [5], breaking the transmission distance record for quantum cryptography. To date, most commercially viable QKD systems rely on fiber-optic links limited to 100 to 120 km. At longer distances, random noise degrades the photon stream. Quantum keys cannot travel far over fiber optic lines, and, thus, they can work only between computers directly connected to each other. To overcome the distance limitation problem and to extend the QKD technology into a network of multiple users we present a QKD network model in the next section.

## 3. PROPOSED MODEL FOR QKD NETWORK

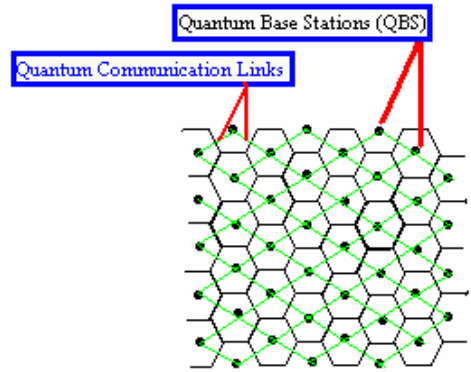

**Figure 4. Quantum Cellular Network**

We present the idea of a quantum cell which is a basic building block of our simple QKD network. Just like mobile cellular system this network consists of cells, see figure 4. The quantum network cell consists of a number of Quantum Network Clients (QNC) and one quantum base station (QBS). The QBS acts as a central junction for all the QNCs. The communication link between the QBSs of adjacent cells is a single mode optical fiber of approximately 100 km. QBS also acts as a gateway for the clients located in the other quantum network cells. The QBS can be thought of as a simple quantum relay, see figure 5.

The idea resembles, in some way, to the LAN enclaves introduced in [4]. The QBS performs the function of an eavesdropper whose strategy is the well known intercept-resend attack [7]. The QBS acts as a trusted member of the quantum cellular network. The fact which must be kept in mind is that the use of the trusted QBS will result in low key generation rate, which imposes some limitations on the number of QBS's in the whole network. The whole quantum cellular network is designed without any far reaching technology like quantum repeaters using entanglement swapping [8] or teleportation [9].

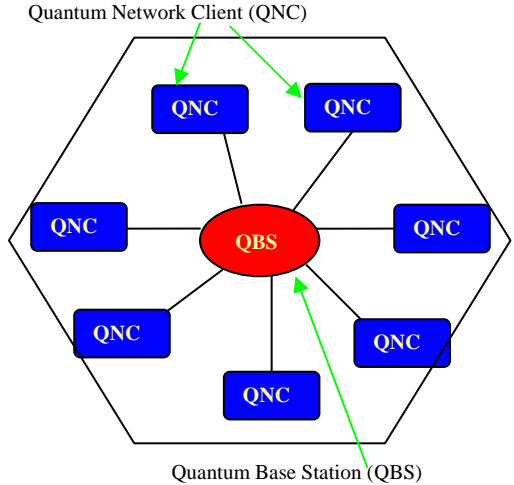

**Figure 5. A Single Quantum Network Cell**

### 3.1. Protocol for the Quantum Cellular Network

We present a protocol which may be thought of as an extension of the well known BB84 protocol [2]. The two QNCs need to share a secret key via the QBS. The QBS is equipped with a quantum random number generator. Steps of the protocol are as follows.

Step 1: The two QNCs who wish to share a secret key inform QBS through classical communication channel.
Step 2: The QBS generates a random sequence of qubits, by using quantum random number generator (QRNG), called the *raw key*, and sends it to both the QNCs who want to share the secret key through quantum communication channel.
Step 3: Each QNC measures each qubit randomly in one of the two measurement basis. The probability that the two QNCs used the same basis for each qubit is 50%.
Step 4: The two QNCs use the classical communication channel to tell QBS about their basis of measurement for each qubit. The probability that the three parties used the same basis of measurement is now 25%.
Step 5: The QBS informs each QNC through classical communication channel about the qubits on which the three parties are agreed. So that sifting can be accomplished.
Step 6: Finally, the error correction and privacy amplification process is done between the two QNCs.

The point of interest, at this moment, is that the keys generated between the QBS and each of the QNCs, say $QNC_1$ and $QNC_2$, are different. Since there are some sets of basis on which the two of the three parties are agreed. There is also a possibility of a set of basis on which all the three parties do not agree. See Table 2.

**Table 2. Creation of Different Keys with Respect to Different Measurement Basis**

|   | $QNC_1$ | QBS | $QNC_2$ |   |
|---|---|---|---|---|
| 1 | $\sigma_x$ | $\sigma_x$ | $\sigma_x$ | Secret Key |
| 2 | $\sigma_x$ | $\sigma_x$ | $\sigma_y$ | Partial Secret key |
| 3 | $\sigma_x$ | $\sigma_y$ | $\sigma_x$ | No secret key |
| 4 | $\sigma_x$ | $\sigma_y$ | $\sigma_y$ | Partial Secret key |
| 5 | $\sigma_y$ | $\sigma_y$ | $\sigma_y$ | Secret Key |
| 6 | $\sigma_y$ | $\sigma_y$ | $\sigma_x$ | Partial Secret key |
| 7 | $\sigma_y$ | $\sigma_x$ | $\sigma_y$ | No secret key |
| 8 | $\sigma_y$ | $\sigma_x$ | $\sigma_x$ | Partial Secret key |

As shown in the above protocol that QBS is released after the process of sifting and is not involved in the error-correction and privacy amplification phase. But one thing which must be taken into account is that the QBS contains the same raw qubits which QNCs posses and QBS has the full access to the public channel which QNCs are using for error-correction and privacy amplification phase. Hence there is a large possibility that QBS finds the secret key which is very similar, but not completely identical, to the one shared by $QNC_1$ and $QNC_2$. But for a good estimation we assume that the QBS is so efficient that it can calculate the same secret key shared by $QNC_1$ and $QNC_2$. Hence the QBS must be considered as a **trusted member** of the whole process.

### 3.2. Distance Extension & Inter-Cell Communication

In view of today's technology, QKD is possible to a maximum range of about 100km due to the losses in the fiber [1]. Consider the situation in which the two parties who wish to share a secret key are located far apart that the distance between them is about more than 200km. In our quantum cellular network we must place these parties in different cells with multiple QBSs involved.

It seems apparently that the size of the secret key decreases with the number of parties involved in the QKD process increases. As in the standard BB84 protocol 50% of the raw qubits are discarded and the other 50% which is a sifted key is used for extracting the actual secret key. But in the presence of our QBS, the sifted key would be the only 25% of the total raw key or the probability of having the same basis among three parties is ¼. The length of the sifted key is inversely proportional to the number of QBS's involved. Hence for n QBS's the probability of having the same measurement basis among all parties is $1/2^{n+1}$.

The solution of this problem is an improved protocol in which the two QNCs i.e. ($QNC_1$ and $QNC_N$) share a secret key passing through N QBSs or N Cells. The protocol is described as follows:

Step 1: The two QNCs who wish to share a secret key inform their respective QBS's i.e. ($QBS_1$ and $QBS_N$) through classical communication channel.
Step 2: All the QBS's which are equipped with the QRNG generate a random secret key and share that key with their next QBS by means of standard BB84 protocol through quantum communication channels.
Step 3: $QBS_1$ sends a random sequence of raw qubits to $QNC_1$ and calculates its XOR with the key shared between $QBS_1$ and its next QBS say $QBS_2$.
Step 4: $QBS_1$ sends the XORed raw key to $QBS_2$ through classical channel.
Step 5: Step 4 is repeated until it reaches $QBS_N$.
Step 6: $QBS_N$ prepares the raw qubits with the received XORed key and sends the original raw qubits to $QNC_N$ through quantum communication link, so that both the $QNC_1$ and $QNC_N$ receive the same raw qubits.
Step 7: Sifting is accomplished between the two QNCs.
Step 8: Finally, the error correction and privacy amplification process is accomplished between $QNC_1$ and $QNC_N$.

The advantage of this protocol is that the final secret key does not shrink with the increasing number of QBSs. In the whole process it is assumed that our QBS has enough intelligence to route the whole cellular network. Details will be studied in the proceeding research work.

## 4. CONCLUSIONS

In this paper we tried to introduce a simple quantum communication network model. This model may be implemented using the technology, available today. The purpose of our proposed network model is to exploit the available QKD technology in the practical networks on large scale.

We have provided the solution of two problems of QKD networks. First is the distance limitation between two parties, sharing a secret quantum key. We solve this problem by introducing the trusted quantum base stations (QBS) in our network model. Second limitation of the QKD technology is that it has not been realized in a network involving a large number of Alices and Bobs. Our network model overcomes this problem by providing the cell based network design.

There are various issues pertaining to our network design like routing solutions among various QBSs and eavesdropping etc. which would be studied in the future research.

## ACKNOWLEDGEMENT

We appreciate the valuable support of Dr. Grégoire Ribordy, the CEO of idQuantique, one of the leading companies of the world in the field of quantum cryptography.